\documentclass[a4paper,11pt]{article}
\pdfoutput=1 

\usepackage{jcappub} 

\usepackage[T1]{fontenc} 

\title{\boldmath Challenges and prospects for better measurements of the CMB intensity spectrum  \note{{\it in extenso} version of the invited talk presented at CMB@50, Princeton 9-12 June 2015}}

\author[a,1,2]{Giorgio Sironi\note{retired since Nov.11th, 2011}}

\affiliation[a]{Physics Department, University of Milano Bicocca \\Piazza della Scienza 3, Milano, Italy}
\emailAdd{giorgio.sironi@unimb.it}
\arxivnumber{:1603.07565 [astro-ph.IM]}

\abstract{Spectral distortions of the Cosmic Microwave Background (CMB) offer the possibility of
probing processes which occurred during the evolution of our Universe going back up to Z$\simeq 10^7$.
Unfortunately all the attempts so far carried out for detecting distortions failed. All of them were based on
comparisons among absolute measurements of the CMB temperature at different frequencies.
We suggest a different approach:
measurements of the frequency derivative of the CMB temperature over large
frequency intervals instead of observations of the absolute temperature at few, well separated,
frequencies as frequently done in
the past, and, direct measurements of the foregrounds which hinder observations, at the same site and with the same radiometer prepared for the search of CMB distortions.  We discuss therefore the perspectives of
new observations in the next years from the ground, at very special sites, or in space as independent
missions or part of other CMB projects}

\begin{document}
\maketitle
\flushbottom
\section{Introduction}
\label{sec:intro}
\par\noindent Observations aimed at checking the blackbody shape of the CMB spectrum
began immediately after the discovery of the CMB by Penzias and Wilson in 1964 \cite{pnz}, but $\sim 25$ years of efforts were necessary to get sure evidence \cite{firas1}.
\par\noindent
According to the most recent evaluations \cite{fix} the blackbody spectrum which best fits the CMB spectrum has temperature :

\begin{equation} \label{TCMB}
T_{CMB} = (2.72548 ~\pm ~ 0.00057) ~K
\end{equation}
\par\noindent and maximum of brightness (per unit frequency interval) at:
\begin{equation}\label{numax}
\nu_B ~=~(58.83~T_{CMB})~=~160.3 ~ GHz
\end{equation}

\par\noindent Deviations from a perfect planckian spectrum, the so called {\it spectral distortions}, are
however expected. Produced by injection in the matter radiation mixture of our Universe of  energy released
by a variety of phenomena which accompany the Universe evolution,  they can be used to probe occurrence
and properties of these phenomena, going back in time up to $Z\simeq 10^7$, about a week  after the Big
Bang, well before the decoupling era at $Z\simeq 10^3$, about 300000 years later, when the CMB spectral
shape and spatial anisotropies were frozen and got the shape we see today.
\par\noindent But other 25 years elapsed and today, 50 years after the discovery of the CMB, no distortion
has been yet discovered, in spite of many
attempts and firm belief by the scientific community that distortions must be present.

\par\noindent Origin and shape of distortions are widely discussed in literature (see next section for references), so in the following we give just a brief summary of what we can expect to see. Then we look at the values of the absolute temperature of the CMB measured at different frequencies: comparisons of values obtained at different frequencies is the only method so far used for searching CMB spectral distortions. But all the attempts failed. We suggest therefore a different observational approach: differential measurements of temperature, with continuous frequency coverage and measurements, by the same system in the same area of sky, of the foregrounds which hinder observation of the CMB features. We finally discuss the perspectives of carrying such measurements.

\section{Expected distortions}
Distortions are triggered whenever the equilibrium of the matter radiation mixture of our Universe is
disturbed by energy injections caused by a variety of phenomena, among them particle annihilations, photon
and particle production, shock wave dissipation, line emissions etc.. Matter - radiation interactions then
smear the original distortions and gradually redistribute the energy  over the CMB spectrum (for a complete
discussion see for instance Partridge \cite{part} and references therein).
\par\noindent Let's call $Z_{inj}$ the epoch when the energy injection occurred and define
\begin{equation}\label{distor}
\delta T(\nu) = T(\nu) - T_{CMB}
\end{equation}
\par\noindent {\it spectral distortion},
\begin{equation}\label{amplidis}
A(\nu) = \frac{\delta T(\nu)}{T_{CMB}} = \frac{T(\nu)}{T_{CMB}} - 1
\end{equation}
\par\noindent  {\it distortion amplitude}, and
\begin{equation}\label{deriv}
\frac{dA}{d\nu} = \frac{1}{T_{CMB}} \frac{dT(\nu)}{d\nu}
\end{equation}
\par\noindent {\it frequency derivative}  of A, where $T_{CMB}$ and $T(\nu)$ are respectively the thermodynamic
tem\-pe\-ra\-ture$^1$  \note{{\it thermodynamic temperature} $T(\nu)$ is the temperature an ideal blackbody
described by the Planck Law should have for producing the brightness $B(\nu)$. It is different from:
i) the {\it brightness temperature} $T_b(\nu) = B(\nu) c^2/2 K \nu^2$, frequently used by radioastronomers,
obtained assuming the Raleigh Jean approximation instead of the Planck Law to link B and T, therefore
\begin{eqnarray}
T(\nu) &=& ~T_b(x_b)~\frac{1}{~\ln(x_b) + 1~} ~~~~~~~~~x_b = h\nu/K T_b     \label{conver1}\\
T_b(\nu) &=&~ T(x)~ \frac{x}{e^x -1} ~~~~~~~~~~~~~~~~~~ x =~ h\nu/KT       \label{conver2}
\end{eqnarray}
\par\noindent ii)the {\it antenna temperature}
$T_a = P/(K \Delta \nu)$ where P is the total power which arrives (from sky and  environment)
to an antenna of effective aperture $A_e$ and total solid angle $\Omega_a$ within a frequency
bandwidth $\Delta \nu$.
Based on Nyquist law, a Raleigh Jeans approximation of
the Planck Law, $T_a$ is the brightness temperature of a signal of average brightness
$\overline{B} = P/(A_e \Omega_a \Delta\nu)$
}
of the CMB undistorted spectrum and the thermodynamic temperature effectively measured at frequency $\nu$.

\par\noindent Evolution and frequency spectrum of the above quantities depend on the phenomenon which disturbed the matter-radiation mixture, the epoch $Z_{inj}$ when the phenomenon occurred and the
processes of matter-radiation
interactions which follow it. These processes are usually  characterized  by different  parameters:
$Y_{ff}$ for free-free transitions, $\mu$, the chemical potential, for the combination of processes which
bring to Bose Einstein instead of Planck Spectra and $y$ for comptonization
(see for instance \cite{part,long} and references therein).

\par\noindent In literature there are many studies (see for instance refs. \cite{zeld,dande,daly,schubla,dezn}) of the distortions we can expect. Carried out solving analitically and/or numerically diffusion equations, they produced a large collection of possible profiles of $A(\nu)$ versus $\nu$ from which it appears that:
\begin{figure}[tbp]
\begin{center}
\hfill
\includegraphics[width=1 \textwidth,origin=c,angle=0]{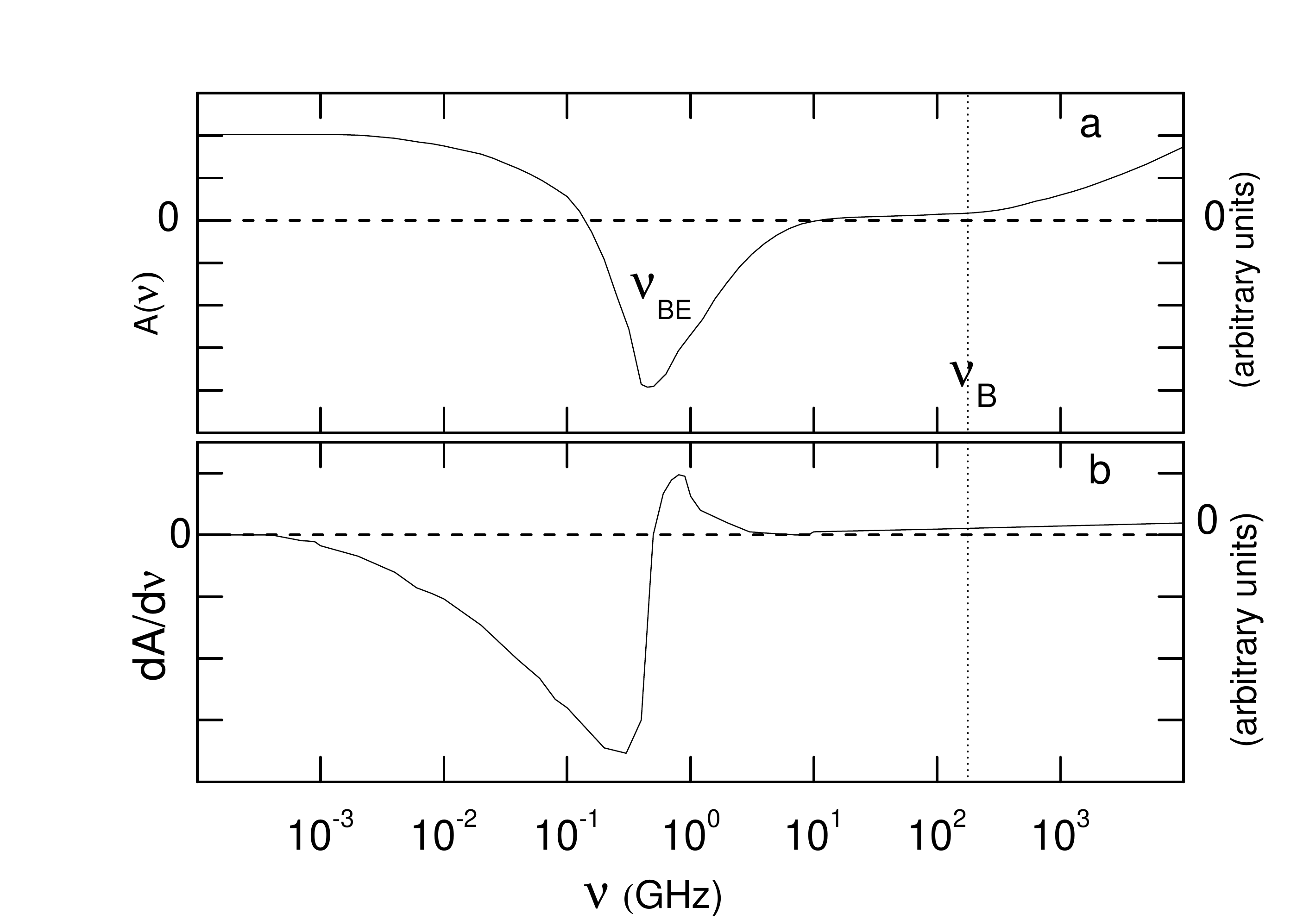}
\caption{\label{fig:1} Trend of the frequency profile of a Bose Einstein distortion produced by an energy injection  $\Delta{\it E}$ at a precise epoch $10^4 < Z_{inj} < 10^7$ . {\it Panel a} : Amplitude A($\nu$), {\it  Panel b} : Frequency derivative of the Amplitude, {\it both Panels} : undistorted spectrum (dashed line). A($\nu$) depends on $Z_{inj}$ and $\Delta{\it E}$, $\nu_{BE}$ on the Universe barion density $\Omega_b$. The effective trend of CMB distortions hopefully we will observe  probably will be a superposition Bose Einstein distortions triggered by different energy injections at different epochs plus distortions produced by energy injections at $Z_{inj} < 10^4$. $\nu_B$ = frequency of maximum CMB brightness.}
\end{center}
\end{figure}
\par i)when $Z_{inj} > Z_{ther}\approx  (10^6 - 10^7)$ matter-radiation interactions are so strong and
fast that thermalization is obtained almost immediately, a new equilibrium spectrum comes out and distortions disappear. When this is the case the distribution of $T(\nu)$ versus $\nu$
is perfectly flat. The only track of energy injections which remains is an increase of the CMB
temperature, but we cannot detect it unless we discover an independent source of information which keeps
track and bring to us information on the previous value of the CMB temperature.
\par ii)when $Z_{inj} < Z_{ther}$  the time elapsed between $Z_{inj}$ and now is insufficient to produce
complete thermalization so footprints of the original distortions survive and should be observable today in
the CMB spectrum.
\par\noindent More precisely
\par a)if $Z_{inj} > Z_c\approx 10^4$ a Bose Einstein semiequilibrium instead of a Planckian equilibrium spectrum is attained, characterized by a {\it chemical potential} $\mu$ proportional to the energy injection $\Delta${\it E} which triggered the distortion. Detailed frequency profiles of the amplitude A($\nu$) of the expected distortions, calculated for different values of $Z_{inj}$ and $\Delta${\it E} can be found in literature  (e.g \cite{dande}- \cite{schubla}). All the Bose Einstein distortions have similar frequency dependence of their amplitudes $A(\nu)$ whose general trend, (see fig. 1, panel a), is characterized  by:
\par -- an almost flat
distribution of $T(\nu)$ versus $\nu$ and minimum distortions ($T(\nu) \simeq T_{CMB}$), between few tens and few hundred GHz, the region around $\nu_B$ (see eq.(\ref{numax}));
\par -- a decrease of $T(\nu)$ below $T_{CMB}$ between few hundred MHz and few ten GHz. As the frequency increases $T(\nu)$ decreases, reaches  a minimum (maximum distortion) at  $\nu = \nu_{BE}$, then goes back toward $T_{CMB}$. The amplitude of the maximum distortion is proportional to $\Delta${\it E} and decreases as $Z_{inj}$ increases; $\nu_{BE}$ depends on $\Omega_b$, the Universe barion density  and slowly increases as $Z_{inj}$ increases. So detecting a Bose Einstein distortion  would be extremely informative ;
\par -- at very low and very high frequencies in many cases $T(\nu)$  goes above $T_{CMB}$.
\par\noindent In the '70, when studies (see for instance \cite{dande,daly}) and searches for distortions
(see section \ref{Tabs}) began,  values of  $|A(\nu_{BE})|$ in the range $10^{-2} - 10^{-3}$  were
not excluded. Today  values of $|A(\nu_{BE})|$ greater than $10^{-4}$ seems improbable and values of $\nu_{BE}\lesssim 1$ GHz are expected.

\par\noindent
\par b)as $Z_{inj}$ goes below $Z_c\approx 10^4$  amplitude, shape and frequency regions  where distortions
should be observable becomes dependent on the process which triggered the distortion and the epoch when it occurred.
There is therefore a variety of expected shapes and amplitudes (see for instance \cite{schubla}).
Among them a special class of distortions characterized by small amplitude and narrow profiles are those
produced at $Z_{inj}\sim 1400$, the recombination epoch, by bound-bound and bound-free helium and hydrogen
transitions.  Usually neglected in the past because the expected amplitudes and profiles are small and narrow
compared to other type of distortions, they are now receiving new attention (\cite{schubla1,apsera}).

\par\noindent \par\noindent The frequency derivative of A($\nu)$ provides detailed information on the shape of $A(\nu)$, for instance it allows to  pinpoint $\nu_{BE}$ in the case of a Bose Einstein distortion or change of slope of $A(\nu)$ profiles. If one day we will arrive to detect distortions, very probably the observed $A(\nu)$ profile will be a superposition of distortions of different types, triggered at different epochs (between $Z\simeq 10^7$ and now), with different values of injected energy. In this case the frequency derivative of the observed profile will help in separating the different type of distortions and recognizing narrow band frequency features combined with wide band frequency features like Bose Einstein distortions.

\section{Search of distortions by absolute measurements of $T(\nu)$}
\label{Tabs}
All the attempts of detecting CMB spectral distortions reported in literature are based on comparisons of absolute values of $T(\nu)$ measured at different frequencies.
\par\noindent Generally (see for instance \cite{trisex,triscmb}) one measures
\begin{equation}\label{DT}
DT(\nu) = T_a(\nu) - T_o(\nu)
\end{equation}
\par\noindent the difference between the radiometer antenna temperature $T_a$ and the brightness temperature $T_o$ of a reference artificial
blackbody $^2$ \note{Usually an optically
thick absorber cooled at the liquid helium temperature,  properly shaped to fill completely the
radiometer beam. Only Wielebinsky \cite{wiele} used an astronomical reference source, the Moon dish, of known
temperature $T_o$.}.
\par\noindent Then, after corrections for the effects (signal attenuation and contamination by system noise)
produced by impedance mismatches of radiometer and reference source, DT is added to $T_o$. This gives the absolute value of $T_a$ from which, after subtraction of $T_{for}$ and $T_{env}$, the brightness temperatures of the signals produced by the sky foregrounds and the radiometer environment (see subsections \ref{forcont}, \ref{envcont} and fig.2) we get the brightness temperature $T_b(\nu)$ of the CMB. Finally $T_b(\nu)$ is converted into  the CMB {\it thermodynamic temperature} $T(\nu)$ and error bars, combinations of statistical and systematics uncertainties of DT, $T_o, T_{for}$ and $T_{env}$, are added to it.

\par\noindent Fig.3, panel a, shows the frequency distribution of a representative sample of measured values of $T(\nu)$
published before 2002 when the majorities of the observations were made (for the complete list see \cite{buri} and  references therein). It includes
results of the coordinated multifrequency observations made by the White Mt (WM) \cite{wm} and South Pole (SP) \cite{sp} collaborations. In panel b of the same figure, results of TRIS \cite{triscmb} and
Arcade \cite{arca}, the two coordinated multifrequency experiments made after 2002, are shown.
\par\noindent With the exception of FIRAS (\cite{firas1,firas2}) and the Gush experiment (\cite{gush}) which cover continuously  extended frequency intervals, the majority of the values of
$T(\nu)$ in literature  are discrete, isolated, single frequency points,
obtained  independently by
different observers, at different sites, using different radiometers. Only after a few
years of observations
groups of observers began to coordinate their efforts, measuring the absolute temperature of the same region of sky,
at different frequencies, from the same site using similar systems
(\cite{wm,sp,trisex,arca} and a common reference source. This way of doing reduced the weight of systematic effects and
improved the accuracy of the final results. A further step forward arrived when observers, instead of using the average properties of the foreground and environment signals, included in their experiment ad hoc observations aimed ad measuring environment and foreground signal (e.g.\cite{part107,trisgal}).
\begin{figure}[tbp]
\centering
\hfill
\includegraphics[width=1 \textwidth,origin=c,angle=0]{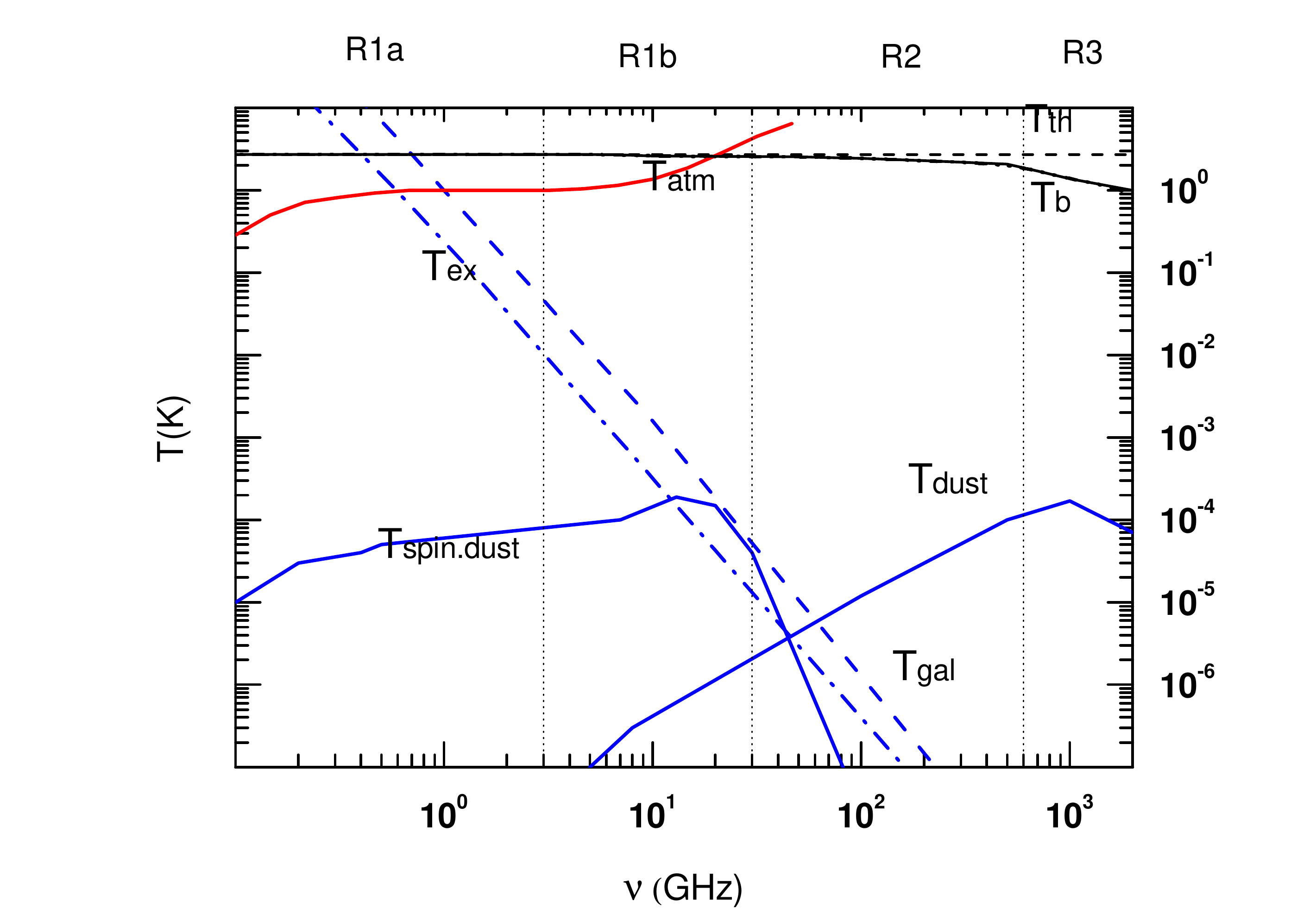}
\caption{\label{fig:2} Brightness temperature of contributions to the antenna temperature.
\centerline {{\it black lines}~~ CMB~~}
 $T_b$ : CMB brightness temperature, ($T_{th}$: CMB thermodynamic temperature, shown for comparison);
\centerline{{\it foregrounds} - blu lines}
$T_{gal}$: galactic free-free and synchrotron; $T_{ex}$: blend of unresolved extragalactic sources; $T_{dust}$: galactic dust; $T_{spin.dust}$ : spinning dust; CO contributions are not shown because they are below $T_{dust}$;
\centerline{{\it environment} - red line}
$T_{atm}$: Earth atmosphere
(at the zenith, 2000 m a.s.l). The ground contribution $T_{ground}$ is highly dependent on ground profile, antenna side- and back-lobes and orientation of the antenna beam therefore it is not shown here.
\centerline{~}
Foreground (with the exception of the extragalactic signal) and environment are anisotropically distributed, therefore their temperatures and spectra depends on the region of sky observed and the radiometer position.  Quoted values and spectra are average quantities.
\centerline{ }
\centerline{R1,R2,R3: frequency regions where methods and conditions of observation are similar(see text)}}
\end{figure}
\begin{figure}[tbp]
\centering 
\hfill
\includegraphics[width=1 \textwidth,origin=c,angle=0]{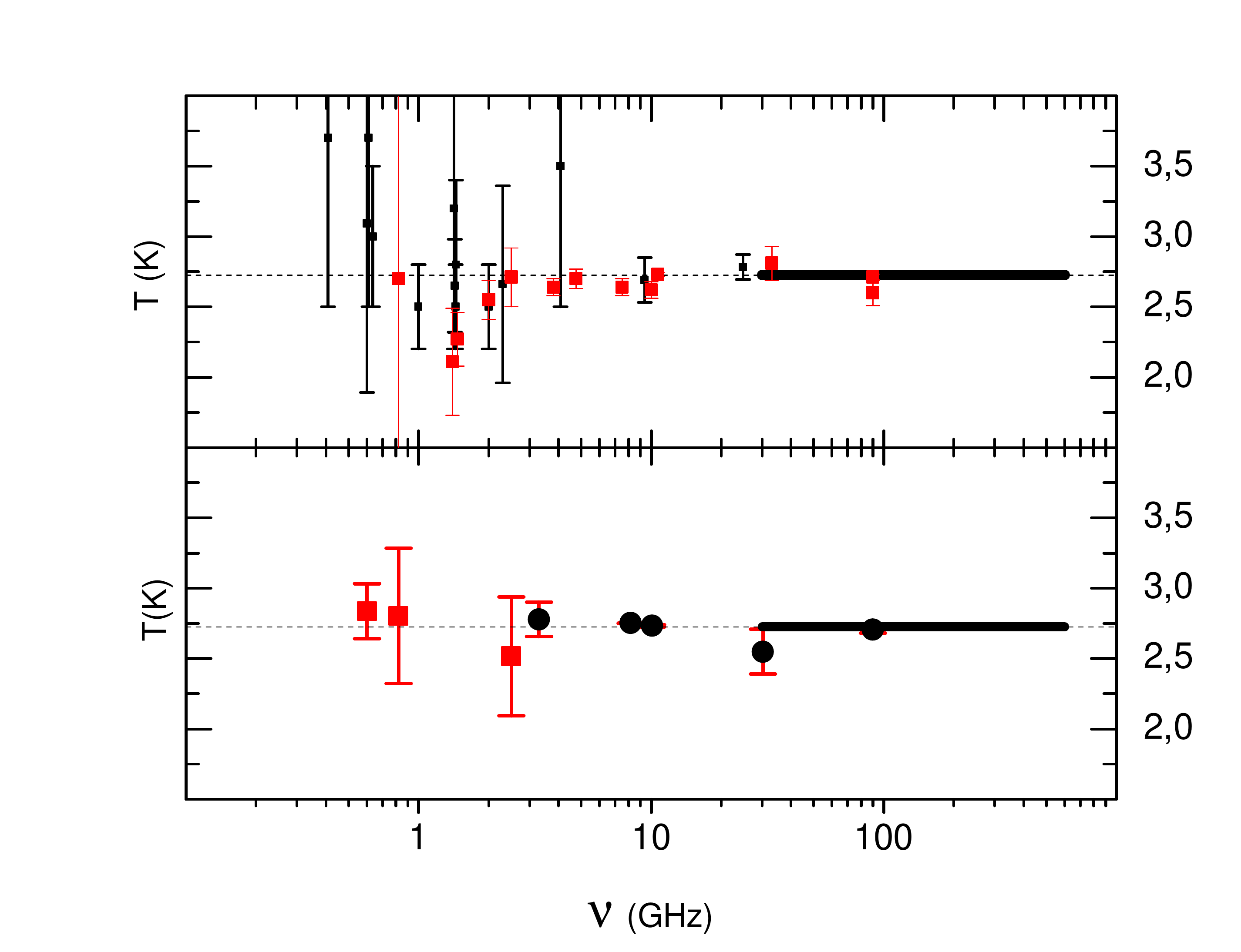}
\caption{\label{fig:3} Collection of measured values of the CMB thermodinamic
temperature.
{\it Panel a}:
old measurements in literature (from the complete list
of Burigana and Salvaterra \cite{buri});  the results of the
coordinated White Mt. \cite{wm} plus South Pole \cite{sp} observations are showen in red;
{\it Panel b}:
recent coordinated
observations: TRIS \cite{triscmb} (red squares) and Arcade2 \cite{arca}
(black dots);
{\it both panels}:
FIRAS final results \cite{firas2} (black horizontal bar) and
best value \cite{fix} of  the undistorted spectrum thermodynamic temperature $T_{CMB}$  (dashed line)}
\end{figure}

\par\noindent From fig.3 and all the data in literature it appears that:
\par i) when the complete set
of data is considered there is no evidence of deviations from a flat distribution;
\par ii)hints, to be confirmed, of deviations can be just glimpsed when results of  coordinated
observations like White Mt \cite{wm}, South Pole \cite{sp},  TRIS \cite{trisex} and Arcade \cite{arca} are
plotted separately: for instance an apparent increase of $T(\nu)$ below 3 GHz (see fig.3, panel b)
has been suggested by Arcade2 \cite{arca}, while an opposite trend, a decrease of $T(\nu)$ with
frequency, seems to be present in the
distributions of the values measured by the WM+SP collaboration. But when error bars are considered both
trends look improbable and completely disappear when the results of all the coordinated experiments are merged. This means that
systematic uncertainties are still present and we have to go further in reducing them.
\par iii)the amplitude of the error bars attached to $T(\nu)$ has a frequency dependence which repeats the
frequency dependence of the foreground components (see fig.2), decreases when results are from coordinated
experiments and is further reduced when observations are made on balloons or in space, where $T_{env}$ is
smaller or negligible.
\par iv) above 600 GHz, in spite of the accuracy of Gush data, results are barely sufficient to confirm that after the peak the brightness spectrum goes down very quickly, as expected in the Wien region of a blackbody spectrum, but we cannot set even poor upper limits to CMB distortions
\par~\par\noindent In conclusion, fifty year after the discovery of the CMB no evidence of
CMB  spectral distortions has been obtained. Only upper limits can be set \cite{triscmb}, ranging
between $\sim 10^{-1}$ K at low
frequencies and $\sim 10^{-3}$ K at high frequency  to the absolute value of the CMB spectral
distortion $|\delta T|$, and

$$ 6~10^{-6} < Y_{ff} < 1.3~10^{-5}$$

$$|\mu_{BE}| < 6~10^{-5} $$

$$|y|  < 1.5~10^{-5}$$

\par\noindent to the parameters which characterize the interaction processes which affect appearance and
evolution of distortions.

\par~\par\noindent Considering the above results, the frequency dependence of
foreground  and CMB  temperatures (see fig.2, 4, 5 and 6)
and the different techniques used by different observers, which gradually go from classical
radioastronomical methods, dominant at frequencies well below $\nu_B$, to IR methods at
$\nu >> \nu_B$, the observations can distributed among three frequency regions:

\subsection{Region 1: low frequency, $(\nu \leq 30 GHz)$}
All the results are single frequency measurements obtained using resonant antennae coupled to low
noise radio
receivers. Belong to this region: i)the first detection of the CMB by Penzias and Wilson \cite{pnz},
ii)the first evidence by  Howell and Shakeshaft \cite{hosh} that the low frequency spectrum of the
just discovered CMB has a $\nu^2$ frequency dependence of the brightness, iii)sets of coordinated measurements made
 at White Mt. \cite{wm}, South Pole \cite{sp}, Campo Imperatore \cite{trisex} and at balloon altitude
\cite{arca}. In region 1 the CMB brightness temperature  is gradually covered by the galactic diffuse emission
when frequency goes below $\sim 1$ GHz and by the atmospheric
signal when goes above 10 GHz (at sea level). Therefore it may be convenient to split
region 1 between subregion 1a - $(0.3 \leq \nu(GHZ) \leq 3)$  and
subregion 1b - $(3 \leq \nu(GHz) \leq 30)$. The region below 0.3 GHz has not been considered and probably will continue to be neglected because here the foregrounds overcome the CMB by orders of magnitude.
\subsection{Region 2: $(30 GHz \leq \nu \leq 600 GHz)$ }
It includes single frequency observations and  measurements with continuous frequency co\-ve1\-ra\-ge made
respectively with antennae coupled to etherodyne systems and radiation concentrators coupled to
bolometric detectors \cite{firas3}.

In region 2 foregrounds are small compared to the CMB while atmospheric effects are so
important that ground based observations would be impossible  were not for two windows at
30 and 90 GHz, used for instance by the White Mt. and South Pole Collaborations.
All the remaining  observations are from stratospheric balloons and in space.

To Region 2 belong also measurements of  $T(\nu)$ at few precise frequencies made studying the
excitation temperature of optical lines of CN and other molecules in interstellar space (see review
by Partridge \cite{part} and references therein).
\subsection{Region 3: high frequency $( \nu > 600 GHz)$}
Here absorption and emission by the earth atmosphere are so large that only observations from
balloons and in space are possible.
\par\noindent Radiometers, very similar to IR and optical systems, with filters to get the radiation spectrum,
can be so small and compact they can be accommodated even in the limited volume available on sounding
rockets (on rockets however the observing time is very short and engine exhausts must be added
to the list of possible contaminants of the sky signal).
In Region 3 observations of the CMB are hindered by the foreground created by dust.

\section{A new approach to the search for spectral distortion}
The FIRAS accuracy ($\sim 5~10^{-3}$~K/(frequency bin)) is the highest so far achieved by absolute measurements but was still insufficient to recognize distortions even integrating over the (30-600 GHz) frequency interval covered by the experiment. Because at FIRAS frequencies  foregrounds are small compared to CMB (see figs.4 and 5) and environment contamination was practically absent because observation were made in space, we conclude that the FIRAS accuracy is entirely set by the quality of radiometer and reference source.
\par\noindent Getting better levels of accuracy for new experiments and extending them outside the FIRAS frequency window means not only improving the evaluation of $T_{for}$ and $T_{env}$, but improving the hardware and reducing error bars, especially systematics among the hardware systems and sub-systems used to carry on the observation, their scales of temperature and the zero levels of these scales.
The elimination of systematics, or at least their drastic reduction, can be immediately obtained if we decide to carry on differential, instead of absolute, measurements of temperature.
\par\noindent The advantages of differential measurements are confirmed by: i)the positive results they brought to the search of CMB anisotropy and polarization where accuracies of few part on $10^6$ are now common and efforts to reach and overcome a few part on $10^9$ are underway (see for instance \cite{planck1} and references therein); ii)the plans of using them also for other type of observations like studies of the 21 cm line in distant galaxies \cite{21line} or the Zeeman effects \cite{zee} or the search of the global signal produced by the blend of the hydrogen  21 cm line in sources at redshift $Z\simeq 6$ (\cite{21-1,21-2})$^3$ \note{the shape of that feature for many  aspects is similar to the shape of the Bose Einstein distortion, but the expected amplitude, (about -100 mK), is higher and occur at about 70 MHz, well below $\nu_{BE}$, in presence of a much higher foreground level.}.
\par\noindent  We  suggest therefore, as we did in the past
\cite{prec}),  that to get better possibilities of detection, new searches for CMB distortions are made:
\par i)measuring the frequency derivative of $T(\nu)$ instead of its absolute temperature;
\par ii)covering continuously the CMB frequency spectrum or large parts of it (It will make easier
recognizing trends and detecting narrow features);
\par iii)including direct measurements of environment and foreground contributions, by ad hoc observations.
\subsection{Measurements of the frequency derivative of the sky brightness temperature} \label{diffmes}
\par\noindent Let's have a differential system which measures the difference of temperature
\begin{eqnarray} \label{varia}
\Delta T_b(\alpha,\delta;a,z;\nu) &=& T_b(\alpha,\delta;a,z;\nu_2) - T_b(\alpha,\delta;a,z;\nu_1) = \nonumber \\
 &=& \Delta T_b^{CMB}(\nu) + \Delta T_{for}(\alpha,\delta;\nu) + \Delta T_{env}(a,z;\nu)
\end{eqnarray}
\par\noindent between the brightness temperatures of two signals collected simultaneously at  frequencies
$\nu_1 = \nu - (\Delta \nu/2)$ and
$\nu_2 = \nu + \Delta \nu/2)$ with $\Delta \nu = (\nu_2 - \nu_1) << \nu = (\nu_1 + \nu_2)/2$ by a
radiometer beam aimed at a point on the sky of celestial and horizon coordinates $(\alpha, \delta)$ and (a,z).

Here $\Delta T_b^{CMB}, \Delta T_{for}$ and $\Delta T_{env}$ are the variations of brightness
temperature of CMB, foregrounds and environment. Subtracting $\Delta T_{for}$  and $\Delta T_{env}$ (see next
section) we get $\Delta T_b^{CMB}(\nu)$ which is then converted into variations of the CMB thermodynamic temperature (see eqs(\ref{conver1}) and
(\ref{conver2} and ref.\cite{dezn}) by
\begin{equation} \label{thvar}
\Delta T(\nu) \simeq \Delta T_b^{CMB}(\nu)~\frac{(e^x -1)^2}{x^2~e^x} ~~~~~~~~ x=h\nu/KT_{CMB},
\end{equation}
\par\noindent   Because (see (\ref{distor}) and(\ref{amplidis}))
\begin{equation} \label{derivapprox}
 \frac{dT(\nu)}{d\nu} = \frac{d}{d\nu}[\delta T(\nu)] =
  T_{CMB}\frac{dA}{d\nu} \simeq \frac{d\Delta T(\nu)}{d\nu}.
\end{equation}\par\noindent  numerical integration of a distribution of measured values of $dT(\nu)/d\nu$ will give

\begin{eqnarray} \label{measdist}
\delta T(\nu) &=& T(\nu) - T_{CMB} =  T_* + \int_{\nu}^{\nu_*}{\frac{\partial{T}}{\partial{\nu}}~d\nu} - T_{CMB}  \simeq \sum_{\nu_*}^{\nu}{\Delta T(\nu_i)}  \\
A(\nu) &\simeq&  \frac{1}{T_{CMB}}\sum_{\nu_*}^{\nu}{\Delta T(\nu_i)}
\end{eqnarray}
\par\noindent where $\nu_*$ is a frequency where (see fig. 1) the distortion amplitude is very close to 0 so
$T(\nu_*) \simeq T_{CMB}$.
\par~\par\noindent
Notice that:
\par i)the differential approach do not improve nor worsen the evaluation of foreground and environment contributions;
\par ii)being the CMB a monopole, the choice of the point on the sky where to look for $\Delta
T(\nu)$ would be independent of $(\alpha, \delta)$ and (a,z) where not for $\Delta T_{for}$  whose
intensity depends on the celestial coordinates of the point where measurements are made and
$\Delta T_{env}(a,z;\nu)$ whose importance depends on the horizon coordinates of the radiometer beam.

\subsection{Contamination from the environment}\label{envcont}
This type of contamination is maximum when observations are made on the Earth or close to it,
(e.g. on balloons). It includes:
\par a)amospheric absorption and emission (see fig. 2). Detailed calculations of $T_{atm}$, the temperature of the atmospheric
signal, its dependence on the elevation of the observing site and the thickness of Earth atmosphere crossed by the radiometer beam and its frequency spectrum are available in
literature (e.g. \cite{aiello} and references therein). They allows to decide immediately if observations at a given frequency can be made from the ground or require stratospheric balloons or space systems.
\par\noindent Because the effective value of $T_{atm}$ depends also on the weather conditions and the position of the observing site to get accurate values, whenever possible,  $T_{atm}$ must be measured, usually by zenith scan methods (e.g. Partridge \cite{part107}).
\par\noindent Having the temperature
of the atmospheric emission the absorption coefficient, necessary to work out the CMB signal corrected
for the absorption by the Earth atmosphere, is immediately obtained
\begin{equation} \label{atmabs}
r = \simeq \frac{T^{phys}_{atm}  - T_{atm}}{T^{phys}_{atm}} \simeq \frac{240 - T_{atm}}{240}
\end{equation}
\par\noindent where $T^{phys}_{atm} \simeq 240 K$ is the average physical temperature of the Earth atmosphere.
\par b)ground emission. Similar to the signal of a  blackbody at $\sim 300$ K which surrounds the radiometer
up to the horizon level, it is collected by the back- and side-lobes of the radiometer beam. The resulting temperature $T_{ground}$ of this signal
depends on: i)ground profile above the horizontal plane, ii)level of
antenna side- and back-lobes, iii)horizontal coordinates of the antenna beam axis. Practically frequency independent, its value can be controlled surrounding the radiation collector with apodized (to reduce diffraction) reflecting screens.
\par c)radiointerferences at various frequencies. By-products of human activities, they arrive from the horizon of the observing site
and from Earth satellites. Once again they can be reduced
by proper sets of screens and frequency filters.
\par d)emission from close astronomical objects, {\it in primis} Sun, Moon and, when observing from space, the Earth.
\par~\par\noindent To get $T_{env}$ and $\Delta T_{env}$ the radiometer beam shape, measured at
a test range paying special attention to the levels of back- and side-lobes, is convolved with the space distribution of the signals from
the ground, close astronomical objects, sources of radiointerferences and atmospheric emission.
\par The importance of $T_{env}$ decreases when observations are made on balloons and go to zero in
space except for radiointerferences and contributions from close astronomical objects, which have to be carefully screened.

\subsection{Foregrounds} \label{forcont}
Foregrounds are a superposition of diffuse emission of galactic and extragalactic origin produced by different mechanisms whose relative importance varies with frequency and the sky regions crossed by the line of sight. The extragalactic component is isotropically distributed, while the distribution of the galactic foregrounds is anisotropic and reflects the underlying structure of the Milky Way.
\par\noindent The most important sources of foregrounds are (see fig.2):
\par 1)Free-free or thermal bremstrahlung process.
\par\noindent Produced by interactions of free electrons with protons and ions of a partially ionized medium, it gives rise to emission from, and absorption in, the region where the electrons and protons move. The brightness temperature of the emission spectrum is characterized by a constant value $T_{ff}$ at $\nu < \nu_o$ and a power law $\nu^{-2.1}$ frequency dependence at $\nu > \nu_o$, $\nu_o$ being set by the densities of free electrons and protons in the medium.
\par\noindent In the interstellar medium $\nu_o \lesssim 1$ MHz, $T_{ff} \simeq 100$ K while absorption of the radiation which crosses it becomes important below few MHz.
\par 2)Synchrotron emission from the interstellar medium.
\par\noindent Produced by the interactions of the cosmic ray electrons with the galactic magnetic field, its brightness temperature has a power law $\nu^{-\alpha}$ frequency  dependence (see fig. 2), with spectral index $\alpha$, (directly measured by radio astronomers and/or extrapolated from the energy spectrum of the cosmic ray electrons, with few per cent error bars), goes from about 2.4 for $0.01 \lesssim \nu \lesssim 0.5$ GHz to 2.8 around 10 GHz and about 3.1 at higher frequencies. Below 10 MHz absorption by the interstellar medium comes in gradually bringing $\alpha$ to zero and negative values below 1 MHz. The variations of $\alpha$ take over smoothly, over frequency intervals whose amplitude and central value depend on the regions of the Milky Way crossed by the line of sight (e.g. \cite{varspeccin, strong,agui,tart}).
\par\noindent Except at frequencies below few MHz where absorption by the interstellar medium begin to appear, the synchrotron signal is always dominant on the free-free signal.
\par\noindent It can be  partially polarized (e.g. \cite{820,mappol} and references therein) and disturbs primarily
observations of the Raleigh Jeans portion of the CMB spectrum.
\par 3)Blend of unresolved extragalactic radio sources. Isotropically distributed, has a power law frequency spectrum similar to, and brightness temperature one or two orders of magnitude weaker than, the galactic synchrotron (see fig.2). It has been studied by various authors (e.g. \cite{blendex} and references therein)
combining the observed counts of the radiosources (the so called log N - log S plots) at various frequencies.
\par 4)Emission from dust present in the interstellar medium, in clouds  and their blend. Its emission spectrum is a modified blackbody spectrum (see fig. 2) at $\sim$ 20 K. It can be partially polarized \cite{pldust,dust} and affects primarily observations in the Wien portion of the CMB spectrum.
\par 5)Emission by spinning dust grains. Its existence has been suggested to explain an apparent increase of the galactic synchrotron  between 20 and 70 GHz, the so called Anomalous Microwave Emission \cite{kogut,ame}.
\par 6)CO line emission associated to dust regions \cite{CO}.
\par~\par\noindent
With the exception of the component of extragalactic origin, frequency spectra, relative weights and spatial distributions of the foreground components and their sum $T_{for}$ depend on the underlying structure of the Galaxy and vary as the radiometer beam moves on the sky. So values and spectra of foreground components shown in fig.2, their sum $T_{for}$ and the the ratio $T_b^{CMB}/T_{for}$ between the brightness temperature of CMB and foreground, shown in fig.4, are not effective but average values. The same is true for $dT_{for}/d\nu$ and $\Delta T_b^{CMB}/\Delta T_{for}$, the ratio between the frequency derivatives of $T_b^{CMB}$ and $T_{for}$, shown in fig.5. The above ratios are sorts of signal (CMB) to noise (foreground) ratios, which can be useful, when we are preparing absolute measurements of temperature, to decide where the CMB is dominant, comparable or buried in the noise, but are insufficient for getting accurate estimates of the contributions to the sky temperature at a given point on the sky.
\par\noindent For measurements of the CMB distortions more convenient quantities are the ratio between the distortion $\delta(\nu) = T_{CMB}(\nu)~A(\nu)$ and the signal $T_{for} + T_b^{CMB}$ which hinders the detection of distortions. It is shown in fig.6 assuming for $A(\nu)$ the amplitude of the Bose Einstein distortion shown in fig.1 and $|\delta(\nu)|_{max}~=~10^{-4}$ K. Fig.7 gives frequency spectra and average values of $\Delta (\delta(\nu))/\Delta (T_{for}+T_b^{CMB})$ and $d(\delta(\nu))/d\nu$. Once again they are average values which can be very different from the effective values we should observe at a given point on the sky, not only because $T_{for}$ is anisotropically distributed, but also because the amplitude $A(\nu)$ of the distortion we hope to detect one day very probably is a superposition of different types of distortions triggered by different energy injections which occurred at different times, with a trend more structured than fig.1 distribution.
\subsection{Extraction of the CMB signal from the sky signal}
\par~\par\noindent  To disentangle the components of $\Delta T_b(\alpha,\delta;a,z;\nu)$ (see eq.(\ref{varia}) and get $\Delta T_b^{CMB}(\nu)$ average values like those plotted in fig 2 and fig.4 to 7 are insufficient.
 Effective values and  precise frequency spectra are necessary. They which can be obtained only from true maps of $\Delta T_b(\nu)$ in a region of sky centered on the point of celestial coordinates $(\alpha,\delta)$ where we want to look for CMB distortions with an accuracy at least comparable to the distortion we want to detect, i.e. $+/- 10^{-4}$ K or better.
\par\noindent In literature there are maps of $T_b(\nu)$, which combined with the frequency dependence of the various components of the sky signal, we can to use for calculating $\Delta T_b$. They cover extended regions of northern and southern sky and were prepared in different years, (by the '60 onward), by different observers, at different ground observatories using different systems (\cite{trisgal} and references therein). But only at a few, low, frequencies it was possible to combine them and prepare maps of the full sky (e.g. \cite{151,408,408r,820,cala}). The situation changed over the last decade when
full sky maps of the foregrounds at high frequencies began to arrive from space experiments \cite{planck}.
Added to the ground observations they now form a large data set which has been used to produce Global Sky Models (GSM) of the diffuse emission of Galactic origin (\cite{21-1,zheng}) and a set of 29 complete maps of the diffuse radiation ($5^o$ angular resolution) distributed between 10 MHz and 5 THz is now available (\cite{zheng}).
But the final accuracy of the temperature measurements is very far from the desired accuracy. According the authors \cite{zheng}, the most recent GSM model has in fact a predictive accuracy ranging between $5\%$ and $15\%$ at most frequencies, $2\%$ accuracy at frequencies where the CMB is dominant, and an overall amplitude offset  $\lesssim 15\%$. Moreover many of the above maps, especially at low frequencies, there include areas which have never been observed. Here structure and temperature distribution have been obtained extrapolating the trend of  the signals measured in nearby areas and/or at other frequencies. Therefore in these regions the maps cannot account nor exclude the existence of local features which may affect the search for CMB distortions.
1\par~\par\noindent So foreground and environment contributions to eq.(\ref{varia}) cannot be obtained from data in literature and/or simple calculations. We have to measure them. The best way of getting them and separating the various contributions to eq.\ref{varia} is:
\par i)mapping $\Delta T_b(\alpha,\delta;a,z;\nu)$ in an area around the point where we want to look for CMB distortions (for measurement of the CMB absolute temperature we should  have to map $T_b(\alpha,\delta;a,z;\nu)$, at frequency intervals regularly distributed inside the frequency interval we want to explore;
\par ii)modelling $\Delta T_b(\alpha,\delta;a,z;\nu)$.
\par\noindent For the foreground a model can be prepared considering the properties of the foregrounds we saw before, or derived from the algorithms of the GSM models in literature, and the known properties and structures of our Galaxy in the mapped area. $\Delta T_{for}$ and the relative weights of the foreground components at the center of the map can be free parameters.
\par\noindent The model of the environment signal, which depends on the structure and position of the observing site and the horizontal coordinates of the radiometer beam, can have $\Delta T_{env}$ and the relative weights of the environment signals can be free parameters;
\par iii)best fitting model and  maps at each frequency will give $\Delta T_b^{CMB}(\nu)$,$\Delta T_{for}(\nu)$ and $\Delta T_{env}(\nu)$;
\par~\par\noindent The computation procedure, far from simple, can be implemented in different ways. They can go from straightforward calculations based on classical methods of bestfitting to very sophisticated methods of model optimization. The choice will depend also on the characteristics of the sky area where observations are going on and the importance of the environment signals. Examples of procedures successfully used for searching weak CMB signal buried in high level of noise like anisotropies at different angular scales or CMB polarization can be found in literature (\cite{jaros,adep}. Liu et a. \cite{21-1} suggest a method for exctracting from the foregrounds the monopole signal produced by the 21 cm line in galaxies at $Z \gtrsim 6$, expected at frequencies close to 70 MHz, with a shape similar to the feature at $\nu_{BE} \simeq  300$ MHz of a CMB Bose Einstein distortion. The Liu procedure offers the possibility of evaluating the sensitivity one can hope to reach when we plan an experiment, but for application to the search of CMB distortions must be adapted to higher frequencies, and wider frequency regions, with a different composition of the fporegrounds (see fig.2).
\par~\par\noindent Whatever the procedure of data reduction one decides to  use the extension of the area to be mapped  must be: \par i)sufficient to provide a number of independent samples large compared to the number of the free parameters we will use to describe foreground and environment signals,
\par ii)sufficiently small to guarantee that in the mapped region the variations of foreground and environment signals are regular and smooth.
\par~\par
\begin{figure}
\includegraphics[width=1 \textwidth,origin=c,angle=0]{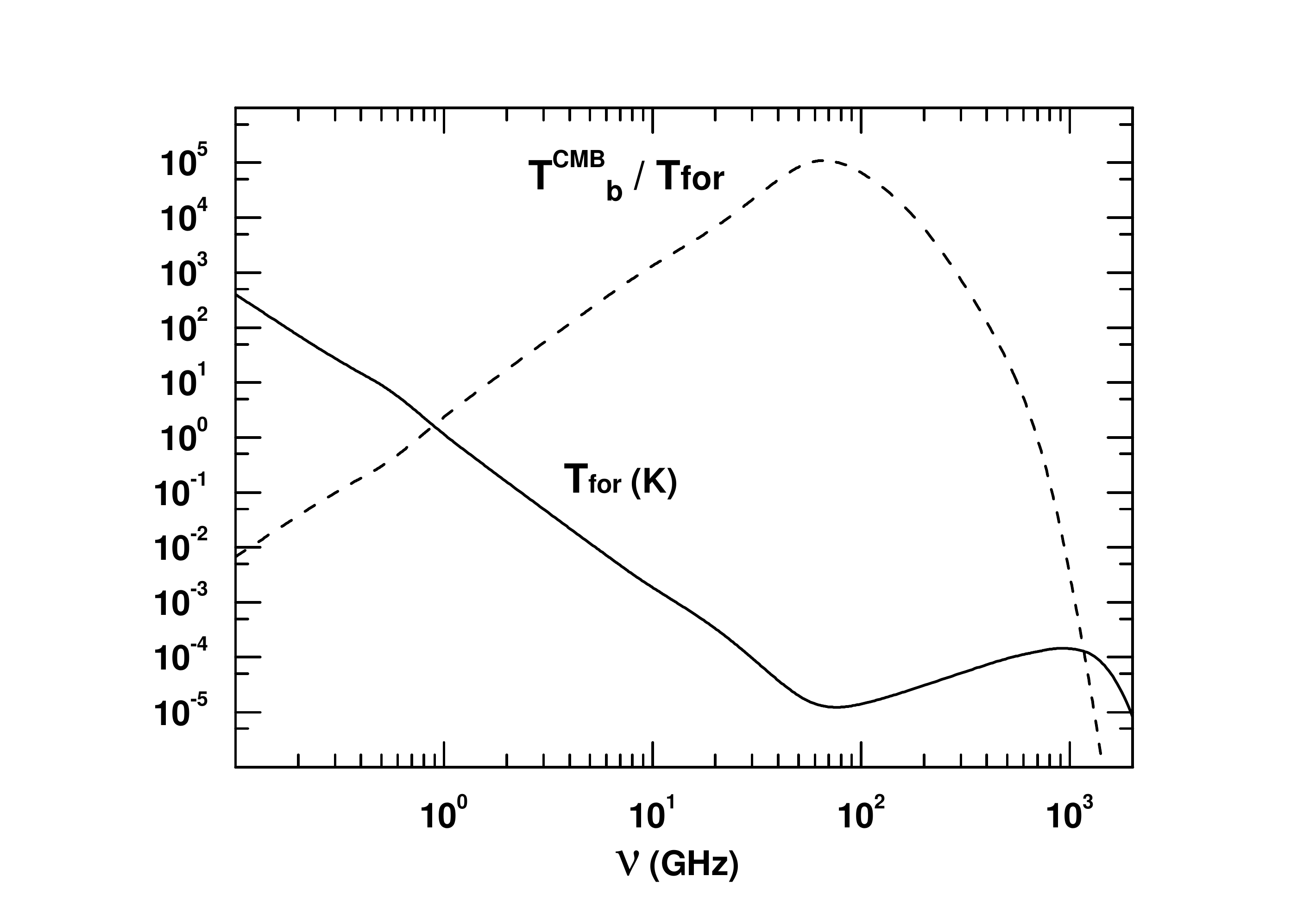}
\caption{\label{fig:4} Average behaviour of $T_{for}$ and $T_b^{CMB}/T_{for}$ frequency spectra and values. Their effective values and spectra depend on the region of sky one is observing}
\end{figure}
\par~\par
\begin{figure}
\includegraphics[width=1 \textwidth,origin=c,angle=0]{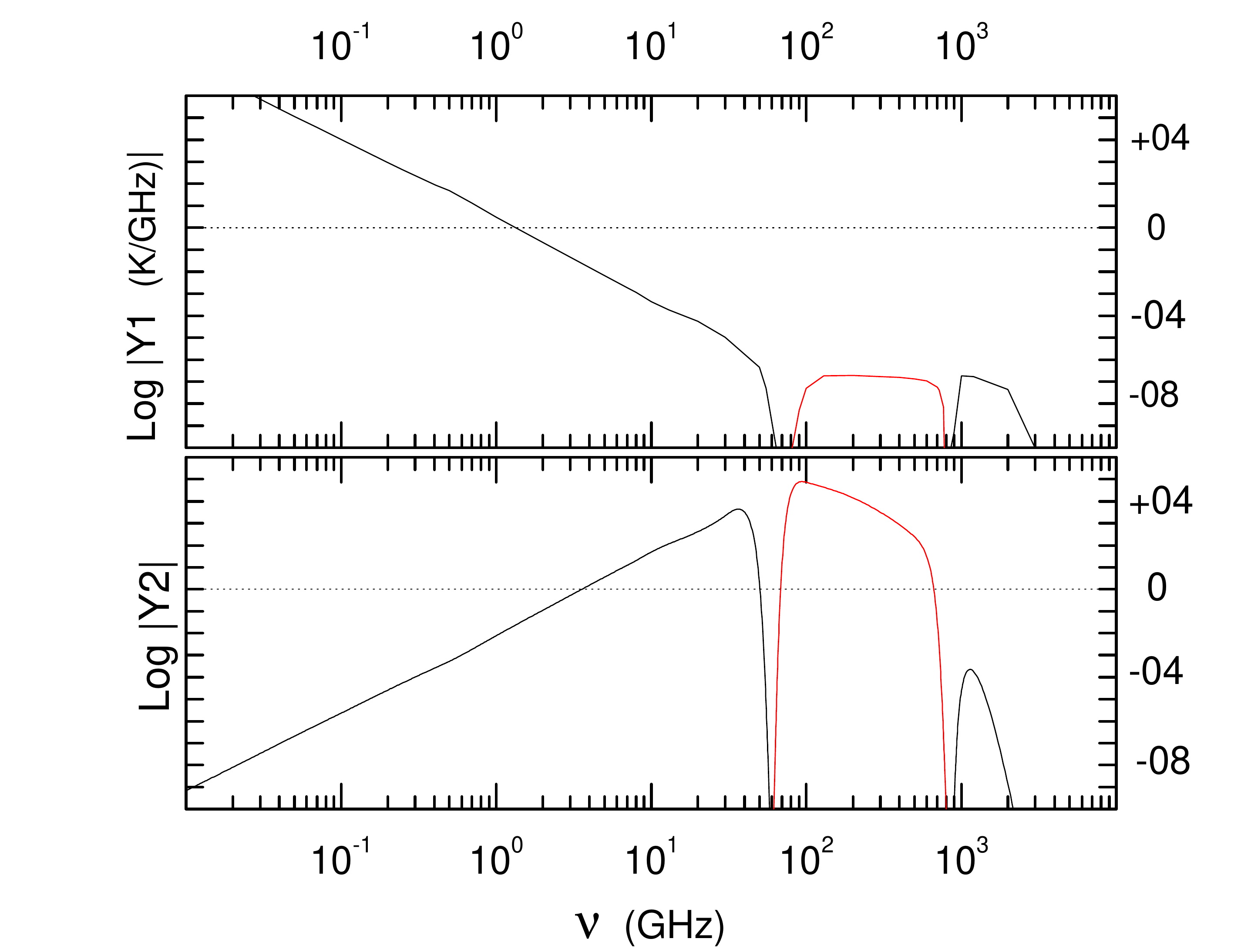}
\caption{\label{fig:5} Average behaviour of $Y1 = dT_{for}/d\nu$ vs $\nu$ (upper panel) and $Y2 = \Delta T_b^{CMB}/\Delta T_{for}$ vs $\nu$ (lower panel). For a better appreciation of the variations of $Y1$ and $Y2$ Log scales have been used and absolute values of Y1 and Y2 plotted (negative values of Y1 and Y2 marked in red). The change of sign of $\Delta T_b^{CMB}/\Delta T_{for}$ around 70 GHz is a consequence of the fact that above that frequency the dust signal overcomes the synchrotron (see fig.2). Plotted values and profiles are typical values. The effective values depend on the observing direction (galactic foregrounds are anisotropically distributed and combine in different ways).}
\end{figure}
\par~\par
\begin{figure}
\includegraphics[width=1 \textwidth,origin=c,angle=0]{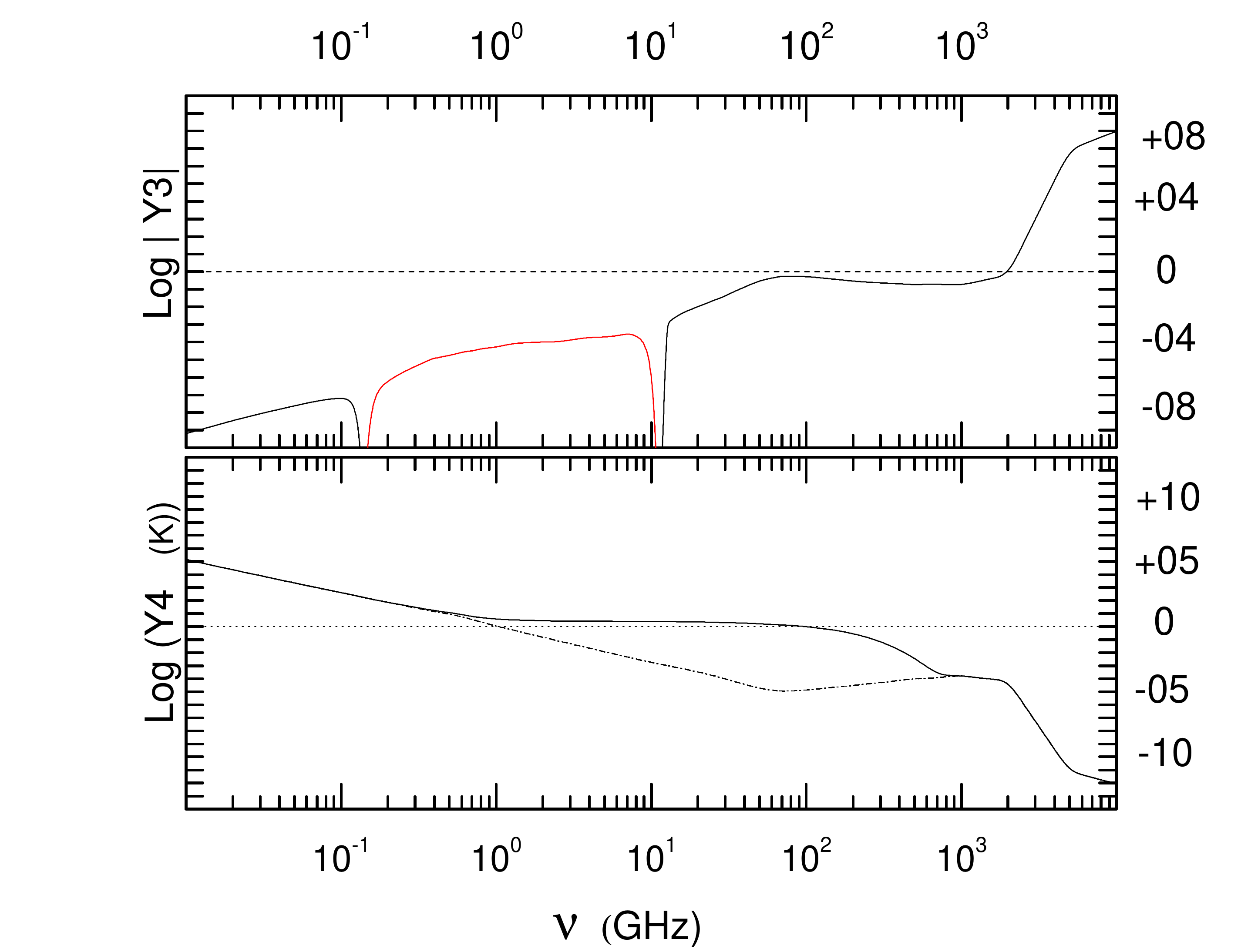}
\caption{\label{fig:6}  Average behaviour of $Y3 =  \delta(\nu)/(T_{for} + T_b^{CMB}) $ vs $\nu$ (upper panel) and $Y4 = T_{for}+T_b^{CMB}$ vs $\nu$ (lower panel) where $\delta$ is the spectral distortion (see eq. 2.1) of a Bose Einstein distortion with the amplitude profile shown in fig.1 and maximum deviation from the undistorted spectrum of $+/- 10^{-4}$ K. For a better appreciation of the variations of $Y3$ and $Y4$ Log scales have been used and absolute values of Y3 and Y4 plotted (negative values of Y3 marked red). The effective values and trend of Y3 and Y4 one can expect to observe at a given site can be more subtle and intricated because the foregrounds are anisotropically distributed and different types of CMB distortions are probably present. }
\end{figure}
\par~\par
\begin{figure}
\includegraphics[width=1 \textwidth,origin=c,angle=0]{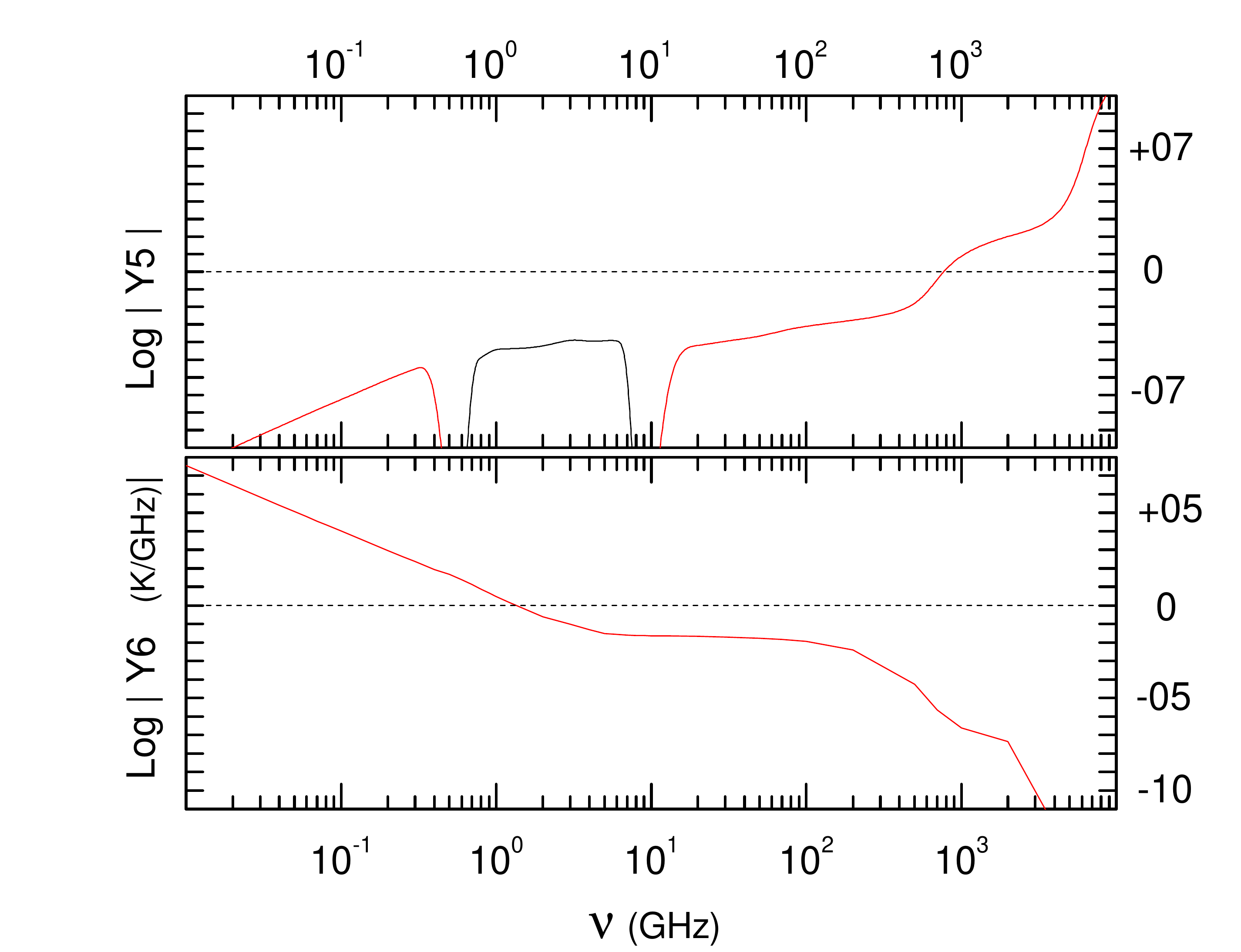}
\caption{\label{fig:7} Average behaviour of $Y5 = |Delta(\delta(\nu)/\Delta(T_{for} + T_b^{CMB}) $ vs $\nu$ (upper panel) and $Y6 = d(T_{for}+T_b^{CMB})/d\nu$ vs $\nu$ (lower panel) for the same Bose Einstein distortion of  fig6. Comments to the caption of fig.6 hold also here}
\end{figure}

\subsection{Additional requirements of future experiments} \label{futurex}
When planning a search for CMB distortions the first step will be looking for sky regions  where foregrounds are minimum and vary in regular way. \par\noindent Then, if observations are from ground observatories additional requirements, important to minimize $T_{env}$, are: radioquiet
site, low horizon profile, high altitude, dry region, latitude close to declination  of the regions of sky to be observed.
\par\noindent When these conditions are satisfied we can can go on  measuring the frequency derivative of $T_b(\nu)$ at regularly distributed  points (separated by $\Delta \theta_o$, the radiometer half power beamwidth, in right ascension and declination) around the point selected for the search of the CMB distortion. The procedure is repeated at different frequencies separated by $\Delta \nu_o$, the radiometer frequency resolution, between the minimum and maximum value of the frequency window to be explored
\par~\par\noindent Other  requirements are:
\par i)radiometer with a well shaped, circular beams, low side- and back lobes (-60 dB), Half Power Beam Width $\Delta \theta_o$.
\par ii)possibility of detecting linear polarization at the 0.1$\%$ level, a foreground signature;
\par\noindent
\par iii)sky area to  be mapped: free of peculiar features, like the North Galactic Spur (\cite{ngsp}) or the Galactic Disk, preferably radio quiet like the region of minimum sky brightness at ($\delta=+35, \alpha=9^h ~30^m$);
\par iv)observing site:
\par\noindent - in space: far from Earth and Sun (e.g. L2 point)
\par\noindent - on Earth ground: at special sites, (high altitude, dry, isolated radioprotected sites), like the Antarctic Plateau during the local winter or the Atacama Desert, for observation of southern sky regions, White Mt. and, at frequencies below few GHz, where atmospheric absorption is small also at low altitudes,  Green Bank, for studies of northern sky regions.
\par\noindent Space and Antarctica are preferable because their environment conditions are particularly stable
and contribute to get low values of $T_{sys}$, the {\it system noise temperature}. The {\it minimum detectable signal}, (1$\sigma$), is in fact
\begin{equation}\label{sensitivity}
 T_{min} = k_s\frac{T_{sys}}{\sqrt{\Delta \nu ~\tau}}
\end{equation}
\par\noindent where $\Delta \nu$ is the system bandwith, $\tau = n \tau_o$ the integration time, $n$ the number of independent observations of the same area of sky, $\tau_o$ the system time constant, $k_s$ a system constant close to 1 whose value depends on the shape of the system bandwidth and the receiver configuration.
\par\noindent If $\Delta \nu = \beta \nu$  ~~eq.(\ref{sensitivity}) gives:
\begin{equation} \label{difsensi}
\Big [\frac{\Delta T}{\delta \nu}\Big ]_{min} \simeq \frac{dT_{min}}{d\nu} = \frac{k_s ~\beta~\tau}{2} \frac{T_{sys}}{(\beta \nu ~\tau)^{3/2}} = \frac{T_{min}}{2 \nu}
\end{equation}

\par v)prefer digital or digitally controlled subsystems (for quick adjustments of the system configuration if unforeseen effects appear);
\par vi)include a reference source, preferably a blackbody of temperature $T_o$, close to the sky temperature. The value of $T_o$ can be poorly known
but its stability must be very high ~($\Delta T/T \leq 10^{-6}$  over a complete observing period);
\par vii) for helping to recognize foreground signals, the capability of measuring linear polarization of the signal is desirable.

\begin{figure}[tbp]
\centering 
\hfill
\includegraphics[width=1 \textwidth,origin=c,angle=0]{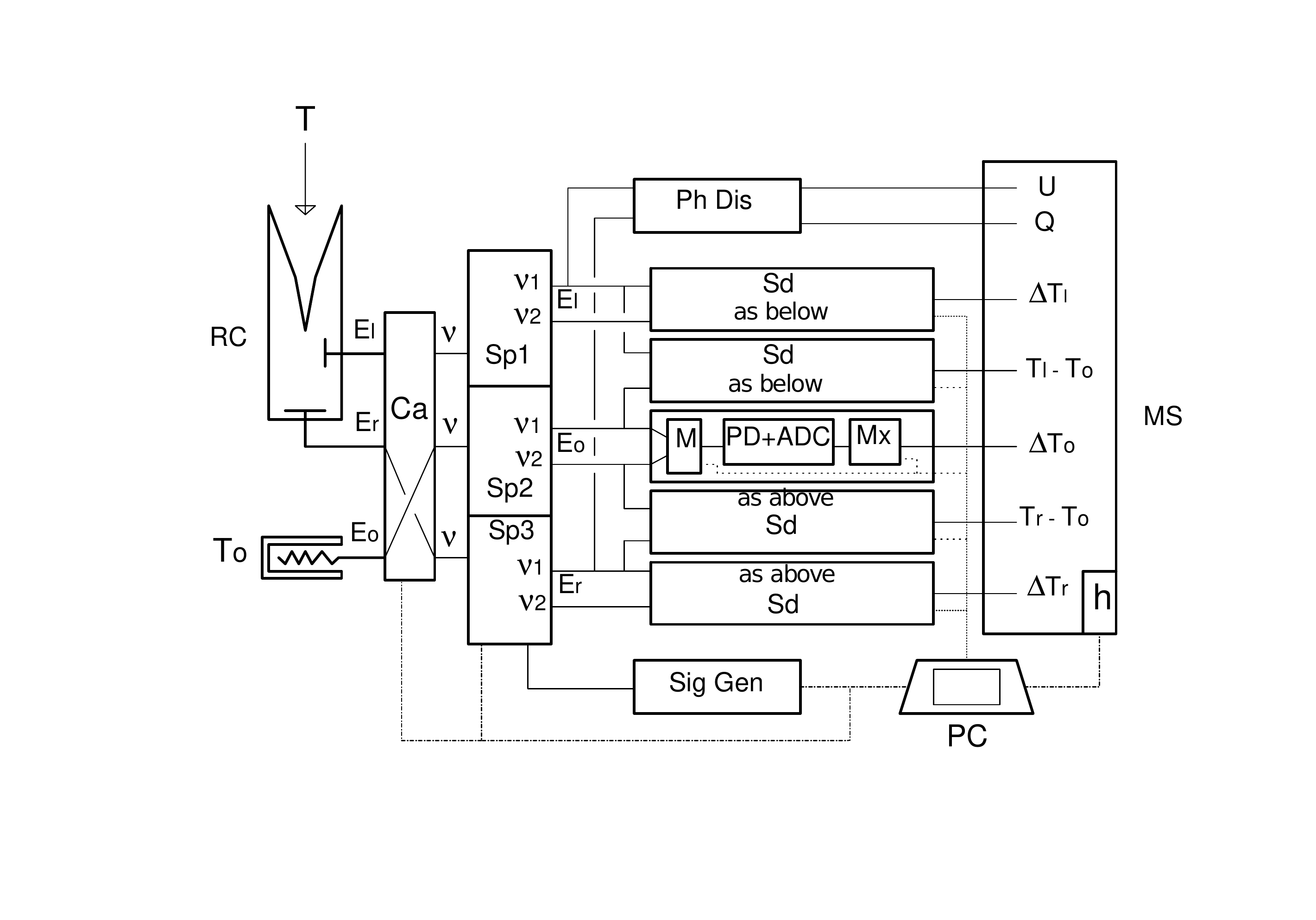}
\caption{\label{fig:8} Block Diagram of a possible configuration of a system for measurements of the CMB spectral
distortions ~~~
T = sky temperature, RC = radiation collector, To= dummy load, Ca = front end multiple switch,
SPi = two channel spectrometer, Sd = synchronous detector, M = modulator, PD = power detector ,
ADC = analog to digital converter, Mx = multiplier, SigGen = high frequency signal generator,
PhDis = phase discriminator, MS = mass storage, PC = system control, El $\&$ Er = signals produced by the left $\&$ right
polarized components of the incoming signal, Eo = dummy load signal, U $\&$ Q = Stokes parameter of the sky signal,
$\Delta T_i$ = difference of temperature  $T_i(\nu_1) - T_i\nu_2)$, h = housekeeping and auxiliary data ($\nu, \nu_1, \nu_2$, time, beam position ...)  .}
\end{figure}

\subsection{Practical system configurations}\label{practic}
Different configurations are possible, for instance the one shown in fig.8. The choice among them depends on the frequency region of
the CMB spectrum to be explored, the technology available and the observing site. Whatever the choice will be
all the configurations must include the following subsystems:

\par i){\it Radiation Collector} (RC).
\par\noindent It must collect the sky signal through a frequency independent, well shaped circular beam, and split it in two orthogonal, linearly polarized components. Beam shape and levels of side- and back- lobes must be carefully
measured before observation, at a level accuracy of least -60dB.  At  high frequencies RC can be a Concentrator (see for instance  \cite{firas3}),
at low frequencies  an under illuminated parabola. In both cases the typical aperture dimension is
\begin{equation}\label{Dimension}
D \simeq 60 \frac{\lambda}{\Delta \theta_o}
\end{equation}
\par\noindent where $\lambda$ is the maximum wavelength
and $\Delta \theta_o$ the system angular resolution ($D\sim 12 \lambda$ for $\theta = 5^o$).
\par~\par\noindent It sets the dimension of the RC aperture. Because the CMB is a monopole the search of CMB distortions would not require a particular beam aperture where not for the foregrounds and environment signal. The great majority of the studies of the CMB spectrum so far made used beamwidths  ranging between few (angular) degrees and $15^o$.
\par\noindent As eq.\ref{Dimension} shows large beams mean antennae/radiation collectors of small aperture.  Usually simple and cheap it can be easily installed in space or at ground observatories, and smooth sky irregularities inside the antenna field of view. But the sky area to be mapped for foreground studies becomes large so finding  regions with regular distribution and minimum values of the foregrounds may be difficult.
.
\par\noindent Small and very small beams on the contrary mean antennae of large apertures, usually  expensive and difficult to install on a space vehicle. They reduce the extension of sky region to be mapped but make the radiometer more sensitive to compact sources and foreground irregularities.
\par For completely new experiments, because we want to observe also at frequencies below 1 GHz, $\Delta \theta_o \simeq 5^o$ is a reasonable trade off but if there are opportunities of using already existing systems beamwidths between $1^o$ and $5^o$ can be accepted.
\par\noindent
\par ii){\it Stable Thermal Reference Source} ($T_o$): a dummy load, temperature stabilized by a bath of boiling liquid with boiling temperature as close as possible to the sky temperature (Helium between 1 and 600 GHz, Argon or Nitrogen at other frequencies). For maximum stability of the load effective temperature the level of the liquid must be kept constant by automatic refilling system;

\par iii){\it Signal Generator} (SigGen)
\par\noindent Digitally controlled set the central frequency of observations and moves it in steps of amplitude $\Delta \nu_o$ over the frequency region to be explored;
\par iv){\it Two Channel Spectrometer} (Sp).
\par\noindent Driven by the Signal Generator  and fed by RC or $T_o$  provides for each frequency $\nu$ two outputs, the sky
signals at $\nu_1 = \nu - \Delta \nu /2$ and $\nu_2 = \nu + \Delta \nu/2$ ~~~( $\Delta \nu/\nu \simeq 0.1$
and $\delta \nu_i/\nu_i \simeq 0.03$) where $\Delta \nu \simeq \Delta \nu_o/3$ and $\delta \nu$ are respectively the frequency separation of the signals
and the spectrometer frequency resolution. Sp can be a
combination of transmission lines and/or omo- or ethero-dyne radiofrequency systems at low frequencies (region
R1), quasi  optical and optical systems at high and very high frequencies (regions R2 and R3);
\par v){\it Synchronous detectors} (Sd). Evolution of the classical Dicke receiver (see for instance \cite{dicke}), includes
Modulator (M), Power Detector (Pd) and Multiplier (Mx)). Its output is the difference  between the
signals at frequency $\nu_1$ and $\nu_2$ which arrive from Sp;
\par vi){\it Power Detector} (PD): cooled bolometer (preferred) or etherodyne receiver (with diodes and low noise preamplifiers),
gives an output proportional to the power associated to the incoming signal;
\par vii){\it Analog to Digital Converter} (ADC): installed as close as possible to the Radiation Collector, for maximum flexibility of system configuration;
\par viii){\it Phase Discriminator} (PhD): evaluates the Stokes Parameters (U and Q or V and Q depending on the
polarization type of the the Radiation Collector outputs, (see for instance MIPOl \cite{mipol},
SPOrt (\cite{sport}) and references therein);
\par ix){\it Multiple Switch} (Ca): for swaping the spectrometer inputs and detecting systematic effects;
\par x){\it Mass Storage} (MS): digital repository of all the measured quantities, plus time,
frequencies $\nu, \nu_1, \nu_2$, system configuration and housekeeping data;
\par xi){\it Control Unit}(PC): drives all the subsystems and data recording system.
\par~\par\noindent Subsystems  ahead of the ADC converters are analogic units digitally controlled by the
system control unit. Subsystems which follow the  ADC converters are digital units
realized via software which can be reconfigured whenever necessary. Therefore closer the ADC units are to the
system front end,  more flexible the system is.
\par~\par\noindent To minimize $T_{sys}$, reduce $T_{min}$ and increase the system sensitivity , RC and front end components must be cooled to a temperature close to the sky temperature.
\par~\par\noindent To check the efficiency of the differential system and its capability of detecting with $1\%$ accuracies distortions of minimum amplitude $10^{-a}$
\par i)the effective temperature $T_o^{eff}$ of the dummy load measured at the input of each of the two channel spectrometers $Sp_i$ must be stable at the level of a part on $10^{(a+2)}$ over the entire observing period and have a known frequency dependence with maximum variation of a part on $10^{(a+2)}$ over $\Delta \nu = \nu_2 -\nu_1$, the frequency separation of the Sp outputs;
\par ii) at each frequency $\nu$ the difference between the two outputs of $Sp_i$, when the spectrometer is fed by the dummy load, must be less than a part on $10^{(a+2)}$.
\par~\par\noindent For instance for  a $10^{-6}$ minimum detectable distorsion,  ($a=6$), the dummy load effective temperature must be stable to a part on $10^8$ while an integration time of 11 day per sky bin will be necessary to allow the detection at 1$\sigma$ c.l. of distortion of absolute amplitude $\geqslant 10^{-6}$.
\par~\par\noindent The absolute value of $T_o^{eff}$ does not affect the measurements of $\Delta T_i$ over the frequency region (e.g. R1 or R2 or R3) explored by one radiometer, but can help to link results obtained by different radiometers and/or observers operating in the same or in a different frequency region.
\par~\par\noindent Calibration of the system response (adc~units/K) can be made (in laboratory and possibly also at the observing site or platform) feeding each $Sp_i$ with the dummy load, and pumping on the bath of the boiling liquid to change its temperature by a known quantity and measuring the variation of the signals at each of the $Sp_i$ outputs.

\section{Perspectives of new searches of CMB spectral distortions}
As already said the differential approach does not improve nor worsen the evaluation of the foreground contributions but makes the accuracies of all the measured quantities independent from the zero level of the scales of temperature and less prone to systematic
effects than absolute measurements. Accuracies of a part on $10^6$ or better are therefore possible.
\par~\par\noindent
To solve completely  the puzzle of the
CMB distortions at least three systems, covering frequency region R1,
R2 and R3 respectively, are necessary (see fig.2):
\par~\par  In Region 1, ((0.3 - 30) GHz),
\par\noindent - below 3 GHz (Region R1a) an underilluminated curved reflector of radius  R = 12 m or more is
necessary, fed by a system of dipoles or by a concentrator, arranged in such a way to produce a frequency independent beam with
$\Delta \theta_o \simeq 5^o$. This will leave on the reflector an unused external rim, $\sim(R - (60/\Delta \theta_o)\lambda)$ which will act as a screen against undesired signals. This reflector aperture makes impossible observations using rockets and balloons and requires deployable systems if we decide for satellites. Sp can be made using classical radioastronomical techniques (combinations of transmission lines and/or omo- or etherodyne systems). Power detection by Diodes fed by low noise amplifiers will be used until good,
cooled, bolometers will become available also at low frequencies.
\par - above 3 GHz (Region R1b) small dimensions of RC and other system components make
observations from stratospheric balloons or even better in space, where environment contamination is practically absent, more appealing. Cooled bolometers can be used for detection.
\par~\par In Region 2 , ((30 - 600) GHz)atmospheric absorption  makes ground observations  impossible, except in two windows
at about 30 and 90 GHz,  and disturbs balloon experiments. Observations therefore must be made in space using satellites which must accommodate radiation collectors with a maximum aperture of 1.2 m and compact subsystems based on microwave and IR techniques. Rockets can be used only for radiometer operating at frequencies $\geqslant 100$ GHz, where the maximum aperture of the radiation collector is 4 cm.
\par\noindent Detection and dispersion systems can be mixtures of microwave and IR techniques.
\par~\par In Region 3 ($\nu > $600 GHz) only rockets and satellites offer a possibility of carrying on observations, accommodating systems more and more compact, similar to  optical radiometers with cooled bolometers for detection and IR derived dispersion systems.

\begin{table}[tbp]
\centering
\begin{tabular}{|l|l|}
\hline
Radiation Collector (RC) & - curved reflector\\
& - diameter 12 m\\
& - steerable along the meridian\\
& - cryogenically cooled concentrator/illuminator \\
& - ground screen\\
\hline
Beam& - frequency independent\\
 & - HPBW = 5$^o$\\
& - side lobes and back lobes $<$ -60 dB\\
\hline
Receiver (Rx)& - triple differential radiometer\\
& - cryogenically cooled between RC and power detectors\\
& - polarization sensitive\\
& - digitally controlled\\
& - $T_{sys} \lesssim 10 K$\\
& - operation frequencies = 0.3 -- 30 GHz\\
\hline
Reference Source ($T_o$) & - criogenically cooled dummy load (see text)\\
& - effective temperature accuracy: 1$\%$\\
& - stability $\Delta T_o/T_o \leqslant 10^{-6}$ during the transit through\\
&~~~~~the antenna beam of a pointlike source\\
\hline
Two Channel spectrometers (Sp)& - frequency separation of the two channels $ \Delta \nu/\nu = 0.1$\\
& - frequency resolution $\delta \nu/\nu = 0.03$\\
\hline
Observation& - at Dome C or Amundsen Scott base (Antarctica)\\
& - transit mode, at nightime\\
& - repeated at different declinations\\
\hline
&\\
Expected sensitivity & - $ [\frac{\Delta T}{\delta \nu}]]_{min} \lesssim 10^{-12}$   ~~K/Hz\\
&\\
\hline
Integration time & - $\leqslant$~~5~~d/sky~bin  ~~~($T_{min} = 5~10^{-6}$ K , ~ 1 $\sigma$)\\
& - $\leqslant$ 1.2~d/sky~bin  ~~~($T_{min} = 10^{-5}$ K , ~~~ 1 $\sigma$)\\
\hline
\end{tabular}
\caption{\label{tab:i} Expected Characteristics of LFGDR (Low Frequency Ground Differential Radiometer) for the search of CMB spectral distortions at low frequencies from Antarctica (see text and fig.3)}
\end{table}

\section{Conclusions}
The best observing conditions can be found in space, especially at distant sites like point L2, where  quiet and stable environment conditions
minimize the system noise and guarantees optimum observing conditions.
But: i)preparing a system for operation in space is very expensive, ii)there are long waiting lists of space
projects, already approved and waiting to be launched or in the approval phase.
1\par~\par1\noindent We cannot reasonably expect
that newly proposed experiments, if accepted, will fly before 2030-2040, so
to be practical and to keep our feet on the ground:
\par - for Regions R2 and R3 we can think about the possibility of becoming part of more ambitious space
and balloons experiments already proposed for measurements of the CMB polarization, among them
PIXIE \cite{pixie}, PRISM \cite{prism}, LSPE \cite{lspe} and their evolutions.
\par\noindent Generally they have narrow beams and plan to explore the whole sky, characteristics which are not necessary for the search of CMB spectral distortions, but useful  to improve our knowledge of the foregrounds. If these experiments will be adapted or already include continuous
frequency coverage and on board evaluation of the frequency derivative of the sky sky signal probably
besides polarization will detect also CMB spectral distortions.
\par - for Region R1 after LOBO \cite{lobomega,lobo,lobo1} and DIMES \cite{dime}  apparently no
other proposals for the search of CMB distortions at  frequencies below few GHz have been submitted to the funding agencies nor can be found
in literature.  Because at these frequencies the dimensions of the radiation collector, ($\sim$12 m for
$5^o$ angular resolution), is definitely too large for balloons  and rockets, ground based observations from the Antarctic Plateau or from the
Atacama desert are the only possibility if we want to carry on observation before the end of the next decade. To operate in the very harsh environment existing there it will be necessary to
get support at the Amundsen Scott base at South Pole (\cite{amund}), or even better at Dome Concordia
Station \cite{concor} for Antarctica or from ALMA \cite{alma} at Atacama. Let's call this experiment LFGDR (Low Frequency Ground Differential Radiometer) (see Table 2).
\par\noindent But if we can wait and ready to plan experiments in Region 1 also from space (let's call it LFSDR, Low Frequency Space Differential Radiometer) deploiable curved reflectors have to be considered.
\par\noindent Studies of deploiable (e.g. \cite{deploy}) and ground based (e.g. \cite{lobomega} ) reflectors
for observation of CMB and diffuse galactic emission  have been made in the past and large reflectors for other
radioastronomical observations have been already sent and deployed in space(e.g. \cite{vsop}).
\par\noindent The need of using a 12 m class reflector will make LFSDR so huge and expensive that, if approved and supported, will probably accommodate as piggy back also the systems for the
search of CMB distortions in regions R2 and R3.

\par~\par Whatever will be the future,  it is necessary to begin immediately ancillary systematic observations
of the foregrounds using already existing radiotelescopes and start technological studies for
improving many radiometer components. Among them:
\par i)optimum geometry of (parabolic or spherical) reflectors with small f number, illuminators and concentrators,
\par ii)cooled illuminators/concentrators
\par iii)low noise, low frequency,
bolometers ,
\par iv)tunable, two outputs, frequency spectrometers,
\par v)stable, low temperature noise sources
\par vi)fast ADC
system operating, at the highest frequencies,
\par vii)digitization
of all subsystems at all the frequencies.
\par\noindent A list of operations which will keep us busy for years,with important outcomes also for other type of research, to be carried anyway.

\acknowledgments
I thanks J.Peebles and the organizers of CMB@50 (Princeton, June 2015) for inviting me to give a talk
on perspective of detecting CMB spectral distortions, which triggered me, when I came back, to put it in written form . I am indebted to Massimo Gervasi, Mario Zannoni
and Enrico Pagana for many discussions, and hope they will have the opportunity, I lost because of
retirement, of detecting CMB distortions.

\end{document}